\documentclass[12pt]{iopart}
\usepackage{graphics}
\usepackage[dvips]{color,graphicx}
\usepackage{subfigure}
\usepackage{xspace}	

\newcommand{\gevc}{GeV/$c$\xspace}
\newcommand{\Pt}{\mbox{$p_{\rm T}$}\xspace}
\newcommand{\MtM}{\mbox{$m_{\rm T} - mass$}\xspace}

\newcommand \sqsn{\mbox{$\sqrt{s_{NN}}$}\xspace}

\begin{document}

\title{
  System size, energy and centrality dependence of strange hadron elliptic flow at STAR
}

\author{H. Masui for the STAR Collaboration}

\address{Lawrence Berkeley National Laboratory, Nuclear Science Division, 1 Cyclotron Road, Berkeley, CA 94720, USA}
\ead{HMasui@lbl.gov}

\begin{abstract}
The elliptic flow ($v_2$) pattern in terms of hadron mass and transverse
momentum $p_{\mathrm{T}}$ is qualitatively described for $p_{\mathrm{T}} <$ 2 GeV/$c$ 
by ideal hydrodynamics in Au + Au collisions at RHIC.
In addition, for $p_{\mathrm{T}} = 2 - 6$ GeV/$c$ the measured $v_2$ follow a universal
scaling by the number of quarks explained by quark coalescence/recombination models.
These observations suggest that a partonic collectivity develops in the matter
in early stage of heavy ion collisions.
Centrality as well as system size and energy dependence of the $v_2$ is important
to shed light on the underlying collision dynamics in heavy ion collisions.
We present the measurements of centrality dependence of $v_2$ at
\sqsn = 200 and 62.4 GeV in Au + Au and Cu + Cu collisions
for $K^0_S$, $\phi$, $\Lambda$, $\Xi$ and $\Omega$
at STAR experiment. We focus on the recent Cu + Cu results and
discuss the centrality dependence of $v_2$ as well as the number of quark
scaling as a function of transverse kinetic energy at different system size and energies. 
We also discuss the eccentricity scaled $v_2$ for identified hadrons 
and implications that ideal hydrodynamical limit has not been reached at RHIC.
\end{abstract}

\maketitle

\section{Introduction}

  Elliptic flow is expected to be one of the most sensitive observable 
  to study early stage of heavy ion collisions at Relativistic Heavy 
  Ion Collider (RHIC), see recent review in~\cite{Voloshin:2008dg}. 
  The elliptic flow is defined by the second harmonic Fourier coefficient 
  of azimuthal distribution of produced particles with respect to the reaction plane
  \begin{equation}
    v_2 = \left<\cos{(2\phi - 2\Psi_{\mathrm{RP}})}\right>,
  \end{equation}
  where $\phi$ is the azimuth of particles in the laboratory frame,
  $\Psi_{\mathrm{RP}}$ is the azimuthal angle of reaction plane 
  which is determined by the direction of impact parameter and 
  beam axis~\cite{Poskanzer:1998yz}. Brackets denote the average 
  over all particles and events.
  In Au + Au collisions at RHIC, it has been found that heavier hadrons has 
  smaller $v_2$ than lighter hadrons for transverse momentum \Pt $<$ 2 \gevc, 
  which is qualitatively explained by hydrodynamical models~\cite{Huovinen:2001cy}.
  Above \Pt = 2 \gevc, the $v_2$ had a reversed trend where 
  protons and $\Lambda$'s have larger $v_2$ than the charged mesons~\cite{Adams:2003am,Adler:2003kt}.
  The reversed trend can be well described by quark recombination/coalescence 
  models~\cite{Molnar:2003ff,Hwa:2003bn,Fries:2003vb} that 
  indicate the $v_2$ has already developed during the 
  early partonic stage prior to the hadronization. 
  It is important to study the systematics of $v_2$, such as 
  system size, beam energy as well as centrality dependence, 
  in order to understand the underlying collision dynamics 
  in heavy ion collisions.

\section{Data Analysis}

  In this analysis, we used minimum bias 24 M events in Cu + Cu collisions 
  at \sqsn = 200 GeV taken at STAR experiment.
  The event centrality was determined by the uncorrected charged particle 
  multiplicity measured by the main TPC within pseudorapidity $|\eta| < 0.5$.
  Particle identifications were performed by the energy loss in 
  the main TPC that enable us to separate $K^0_S$, $\Lambda$ and 
  $\Xi$ in 0.2 $<$ \Pt $<$ 4 \gevc. 
  Event plane method~\cite{Poskanzer:1998yz} was used to measure 
  $v_2$ at midrapidity. The $v_2$ can be expressed as
  \begin{equation}
    v_2 = \frac{\left<\cos{(2\phi - 2\Psi_2^{\mathrm{EP}})}\right>}{\left<\cos{(2\Psi_2^{\mathrm{EP}} - \Psi_{\mathrm{RP}})}\right>},
    \label{eq:eq2}
  \end{equation}
  where $\Psi_2^{\mathrm{EP}}$ denote the second harmonic event plane 
  determined from elliptic flow, and brackets represent the average 
  over detected particles and events. 
  Event plane was determined by using the forward time projection 
  chamber (FTPC) covered at 2.5 $< |\eta| < $ 4.
  Since the FTPC has a rapidity gap from the main TPC region, it can reduce 
  the non-flow contributions that are not correlated with 
  the reaction plane, such as di-jets and resonance decays etc.
  The denominator in (\ref{eq:eq2}) is defined as an event plane 
  resolution that is about 0.18 in midcentral Cu + Cu collisions.
  We also analyzed the data in Cu + Cu collisions at \sqsn = 62.4 GeV,
  which has 12.5 M minimum bias events. The event plane resolution 
  was by a factor of 2 smaller than that at 200 GeV due to lower multiplicity.

\section{Results}

  \begin{figure}[htbp]
    \subfigure[$v_2$ vs $p_{\mathrm{T}}$]{
      \includegraphics[width=0.5\linewidth]{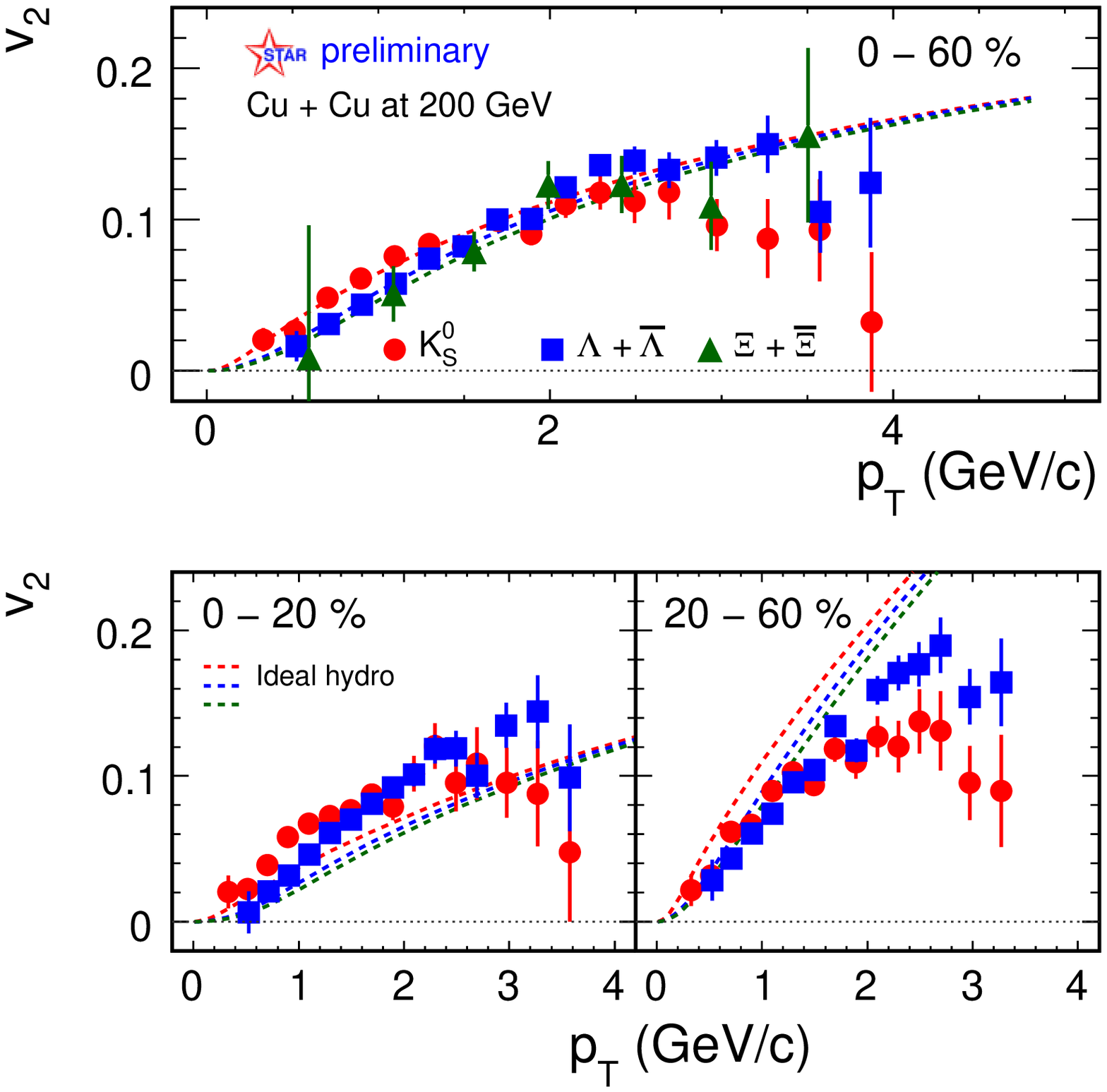}
    }
    \subfigure[$v_2/n_{\mathrm{q}}$ vs $(m_{\mathrm{T}} - mass)/n_{\mathrm{q}}$]{
      \includegraphics[width=0.5\linewidth]{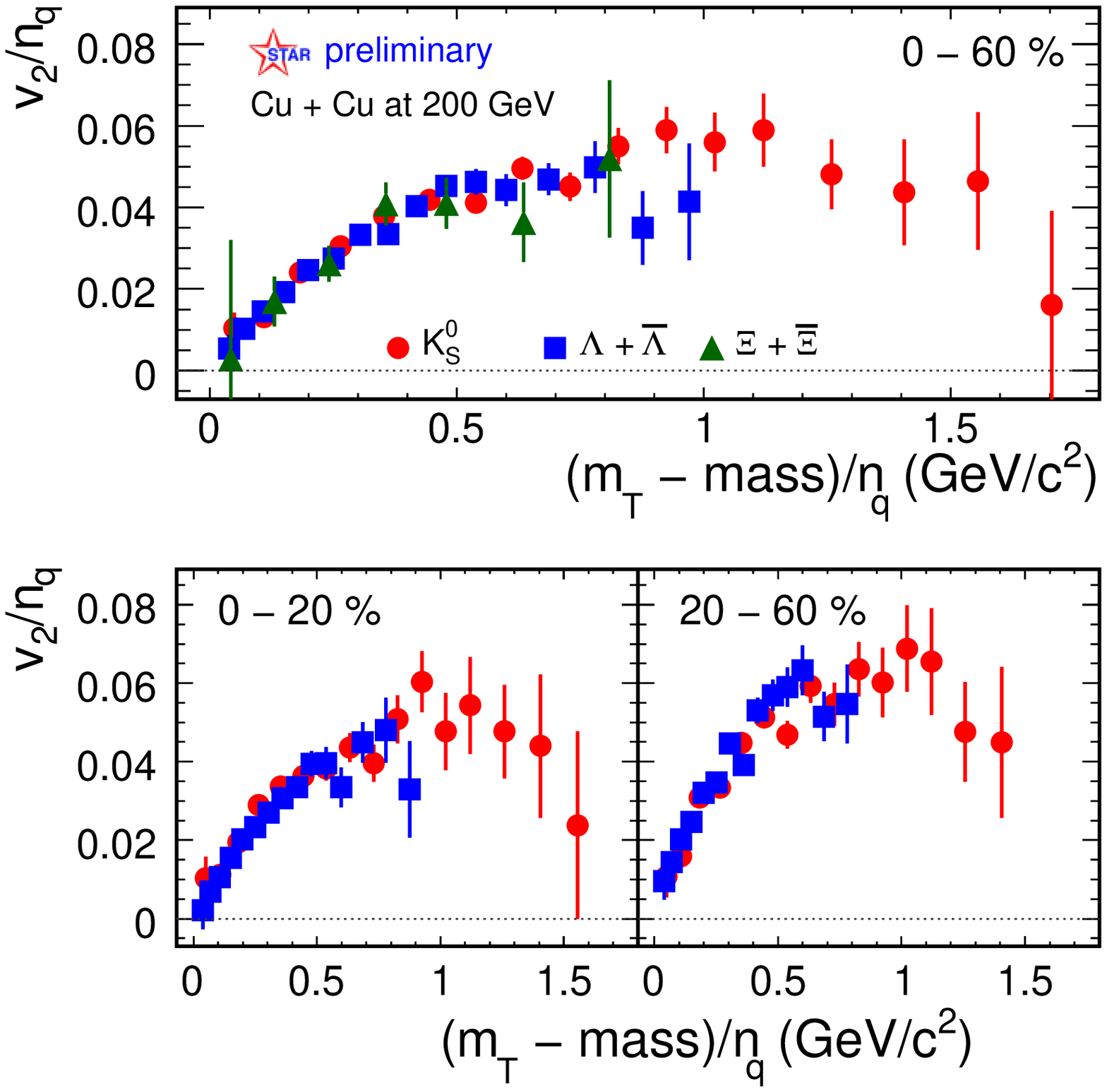}
    }
    \caption{\label{fig:fig01_fig02}
      (a) The $v_2$ as a function of $p_{\mathrm{T}}$ for $K^0_S$, $\Lambda + \bar{\Lambda}$,
      $\Xi + \bar{\Xi}$ in 0 - 60 \% (top), 0 - 20 \% and 20 - 60 \% centralities (bottom).
      Dashed lines represent ideal hydrodynamical calculation~\cite{Huovinen:2006aa}.
      Only statistical errors are shown.
      (b) Number of quark scaling of $v_2$ as a function of \MtM in 0 - 60 \% (top),
      0 - 20 \% and 20 - 60 \% (bottom). Data symbols are the same as those in (a).
    }
  \end{figure}
    Figure \ref{fig:fig01_fig02} (a) shows the $v_2$ for $K^0_S$, $\Lambda + \bar{\Lambda}$ and 
  $\Xi + \bar{\Xi}$ as a function of \Pt in different centrality selections. 
  We have observed that $\Lambda$ has smaller $v_2$ than $K^0_S$ below \Pt = 2 \gevc in 
  Cu + Cu collisions. For \Pt $>$ 2 \gevc, however, $\Lambda$ $v_2$ become larger than 
  that of $K^0_S$. We have also found $\Xi$ has sizable $v_2$ in minimum bias 0 - 60 \% 
  centrality. Results were compared to an ideal hydrodynamical calculation with the first 
  order phase transition from the QGP to the hadron phase. 
  The ideal hydrodynamical model does not describe the centrality dependence of our data.
  For 0 - 20 \%, the model under-predicts our data and for 20 - 60 \%, it over-predicts the $v_2$.
  Effects not included in the model which may be relevant are geometric fluctuations 
  in the initial conditions (particularly important in central collisions) and 
  finite viscosity effects. It remains to be seen if these effects can account 
  for the difference between the models and data.
    Figure \ref{fig:fig01_fig02} (b) shows the number of quark (NQ) scaling of $v_2$ 
  as a function of transverse mass minus particle mass \MtM, namely both 
  vertical and horizontal axes are divided by the number of valence quarks 
  for each hadron. 
  One can see that all measured hadrons lie on the universal curve up to 
  \Pt = 4 \gevc in all centrality bins.
  This results indicate that the $v_2$ has already established in the partonic 
  stage in Cu + Cu collisions.

  \begin{figure}[htbp]
    \includegraphics[width=1.0\linewidth]{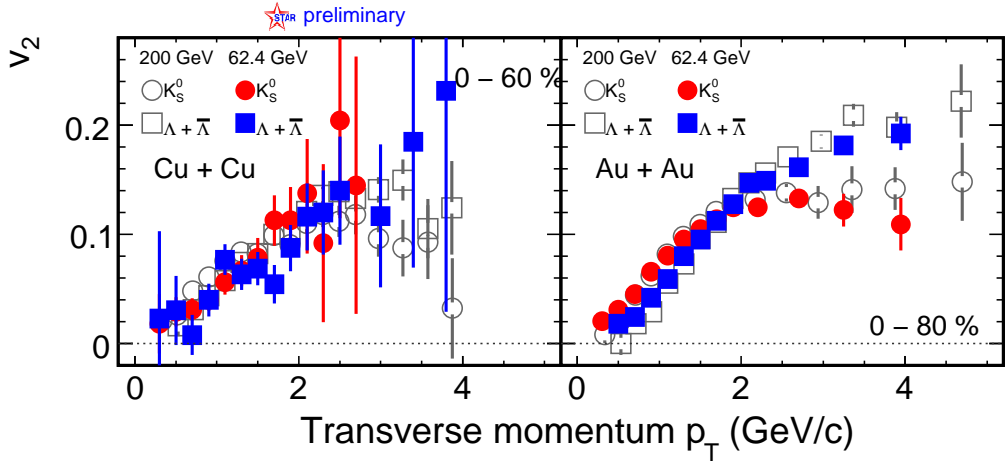}
    \caption{\label{fig:fig03}
      Comparison of $v_2(\Pt)$ for $K^0_S$ and $\Lambda + \bar{\Lambda}$ 
      in 0 - 60 \% centrality
      at \sqsn = 200 GeV with those at 62.4 GeV in Cu + Cu (left)
      and Au + Au collisions (right). The $v_2$ in Au + Au collisions are 
      taken from~\cite{:2008ed,Abelev:2007qg}. Only statistical errors 
      are shown.
    }
  \end{figure}
  \begin{figure}[htbp]
    \begin{minipage}{0.5\hsize}
      \includegraphics[scale=0.5]{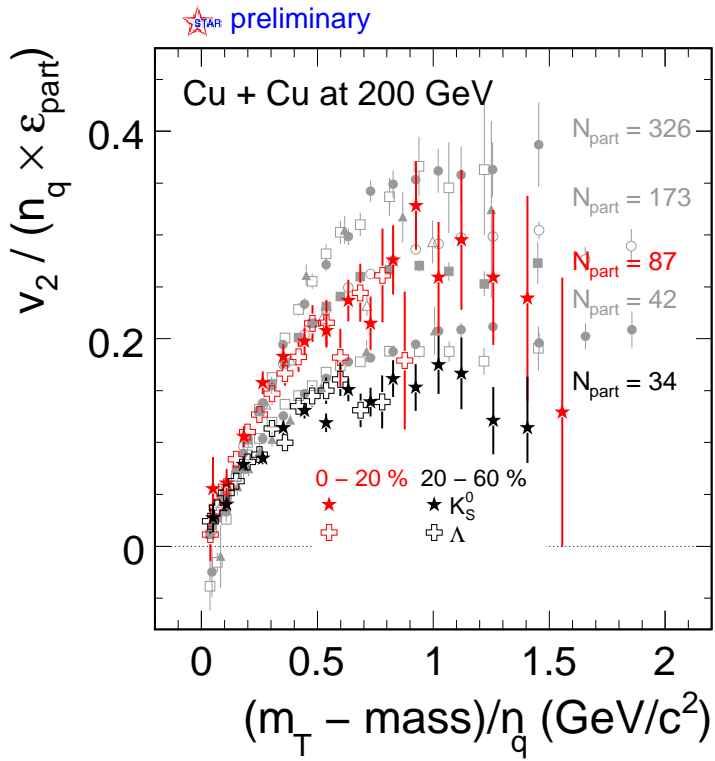}
    \end{minipage}
    \begin{minipage}{0.5\hsize}
      \caption{\label{fig:fig04}
        Number of quark scaling of $v_2$ divided by participant eccentricity 
        $\varepsilon_{part}$ for $K^0_S$ and $\Lambda + \bar{\Lambda}$ 
        as a function of \MtM in Cu + Cu collisions. For comparison, 
        results in Au + Au collisions are also plotted.
      }
    \end{minipage}
  \end{figure}

    Figure \ref{fig:fig03} show the comparison of $v_2$ for $K^0_S$ and $\Lambda + \bar{\Lambda}$
  at \sqsn = 200 and 62.4 GeV in Cu + Cu collisions (left panel).
  For comparison, results in Au + Au collisions are presented in right panel~\cite{:2008ed,Abelev:2007qg}.
  We found that the measured $v_2$ at \sqsn = 200 GeV are consistent with those at \sqsn = 
  62.4 GeV in both Cu + Cu and Au + Au collisions.

    Figure \ref{fig:fig04} shows the eccentricity scaled $v_2$ for 
  $K^0_S$ and $\Lambda + \bar{\Lambda}$ as a function of \MtM. In order to compare 
  different particle species, NQ scaling was also applied.
  Since the event plane at the FTPC was 
  determined by particles from participant nucleons, appropriate initial geometrical 
  anisotropy should be participant eccentricity $\varepsilon_{part}$ that include 
  event-by-event position fluctuation of participants~\cite{Manly:2005zy}.
  The scaled $v_2$ in 0 - 20 \% appeared to be larger than those in 20 - 60 \%, 
  which indicate that stronger 
  collective flow has developed in more central collisions. In addition, scaled 
  $v_2$ in Cu + Cu collisions are consistent with that in Au + Au collisions 
  with similar number of participant $N_{part}$. This result suggest the observed 
  $v_2$ is determined not only by initial geometry, but also by the $N_{part}$.
  Therefore, it is important to understand whether the $v_2$ in 
  most central Au + Au collisions reach ideal hydrodynamical limit or not.
  Because ideal hydrodynamical models predict $v_2$ is only determined by 
  the initial spatial anisotropy.

  \begin{figure}[htbp]
    \includegraphics[width=1.0\linewidth]{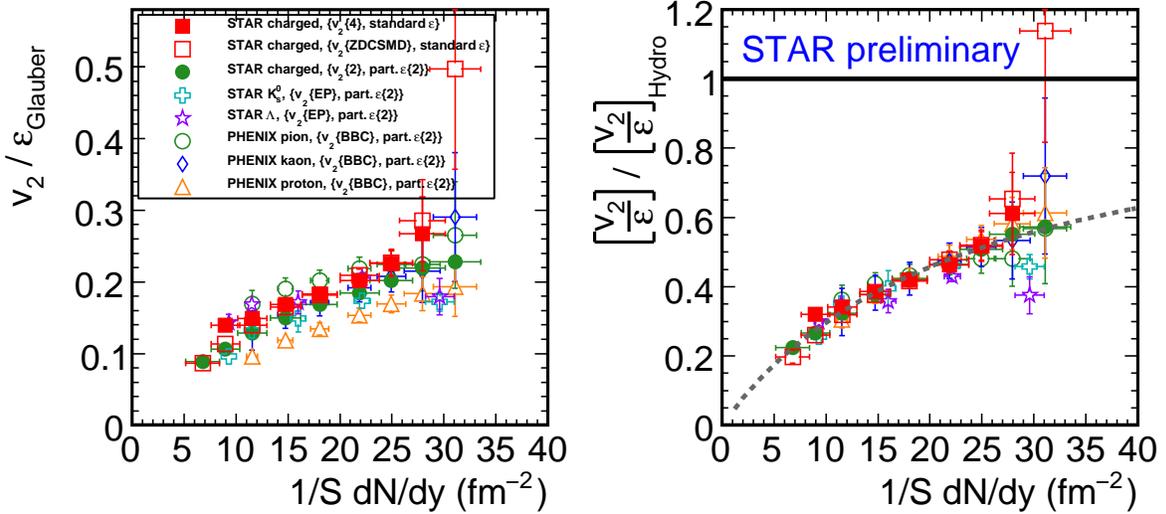}
    \caption{\label{fig:fig05}
      (Left) $v_2/\varepsilon$ as a function of transverse number density $(1/S) dN/dy$
      for different particle species from STAR and PHENIX. STAR results include
      preliminary $v_2$\{4\}, $v_2$\{ZDCSMD\} in Au + Au collisions 
      as well as $v_2$ for $K^0_S$ and $\Lambda$ in Cu + Cu collisions, 
      and published $v_2$\{2\}~\cite{Adams:2004bi}, $v_2$ for $K^0_S$ and $\Lambda$~\cite{Adams:2003am}.
      PHENIX results were taken from~\cite{Issah:2006qn}.
      Error bars denote quadratic sum of statistical and systematic uncertainties.
      (Right) $v_2/\varepsilon$ divided by extracted hydrodynamical limit from the fit 
      as a function of transverse number density $(1/S) dN/dy$. Dashed line represent 
      the simultaneous fit for all measured $v_2$ by (\ref{eq:eq3}).
    }
  \end{figure}
    Figure \ref{fig:fig05} show the comparison of $v_2/\varepsilon$ as a function 
  of transverse number density $(1/S)dN/dy$ for different particle species 
  in Au + Au collisions at \sqsn = 200 GeV from PHENIX and STAR 
  experiments. Published STAR and PHENIX results were taken from~\cite{Adams:2003am,Adams:2004bi,Issah:2006qn}.
  In order to extract the ideal hydrodynamical limit,
  we have used the formula based on a transport model with a finite 
  Knudsen number~\cite{Bhalerao:2005mm}
  \begin{equation}
    \frac{v_2}{\varepsilon} = \left(\frac{v_2}{\varepsilon}\right)_{\mathrm{Hydro}} 
    \left( \frac{1}{1 + K/K_0} \right), \quad \frac{1}{K} = \frac{\sigma c_s}{S} \frac{dN}{dy},
    \label{eq:eq3}
  \end{equation}
  where $\left(v_2/\varepsilon\right)_{\mathrm{Hydro}}$ is the ideal hydrodynamical 
  limit of $v_2$, $K$ is the Knudsen number, $c_s$ is the speed of sound, 
  $\sigma$ is the partonic cross section,
  $S$ is the transverse area of collision zone 
  and $dN/dy$ is the total hadron multiplicity.
  We adopted $K_0$ = 0.7 and $c_s = 1/\sqrt{3}$ in order to reproduce 
  the transport model calculation~\cite{Gombeaud:2007ub}.
  Hydrodynamical limits were extracted for different hadrons
  by fitting the available $v_2$ simultaneously,
  where we assumed $\sigma$ is the same for all hadrons.
  Right panel in Figure~\ref{fig:fig05} shows the ratio of
  $v_2$ to the resulting hydrodynamical limit from the fit 
  as a function of $(1/S)dN/dy$. The ratio should be 1 
  when $v_2$ reach hydrodynamical limit.
  We found that the ratio converge into a universal curve 
  and ideal hydrodynamical limit has not been 
  reached even at the most central collisions at RHIC.

\section{Conclusions}

  In summary, we have measured $v_2$ for $K^0_S$, $\Lambda + \bar{\Lambda}$ and 
  $\Xi + \bar{\Xi}$ at \sqsn = 200 GeV in Cu + Cu collisions.
  We found that $\Lambda$ has smaller $v_2$ than $K^0_S$ below \Pt = 2 \gevc, 
  whereas the trend was reversed above \Pt = 2 \gevc.
  The measured hadrons follow the number of quark scaling as a function of \MtM
  in Cu + Cu collisions at \sqsn = 200 GeV. 
  This result is consistent with the idea that collective flow has been 
  established in the partonic stage before the hadronization takes place.
  The $v_2$ at \sqsn = 62.4 GeV were consistent with those in \sqsn = 200 GeV 
  in Cu + Cu collisions.
  We observed that eccentricity scaled $v_2$ was larger in more 
  central events in Cu + Cu collisions, which indicate that 
  stronger collective flow was developed in more central collisions. 
  Finally, from the comparison of the data with a transport model approach,
  we found that ideal hydrodynamical limit has not been reached 
  even at the most central collisions at RHIC. \\

\bibliographystyle{h-elsevier}

\bibliography{reference}

\end{document}